\documentclass[sigconf,nonacm,9pt]{acmart}

\AtBeginDocument{%
  }

\usepackage{xspace}
\newcommand{\sysname}{\texttt{Flip-Agent}\xspace}

\usepackage{listings}
\usepackage{multirow}
\usepackage[skins]{tcolorbox}
\tcbuselibrary{breakable,skins}

\usepackage{amsmath,amsfonts}
\usepackage{algorithmic}
\usepackage{algorithm}
\usepackage{array}
\usepackage[caption=false,font=normalsize,labelfont=sf,textfont=sf]{subfig}
\usepackage{textcomp}
\usepackage{stfloats}
\usepackage{url}
\usepackage{verbatim}
\usepackage{graphicx}
\usepackage{graphicx}
\usepackage[utf8]{inputenc}
\usepackage{tcolorbox}
\usepackage{amsmath}
\usepackage[table]{xcolor}

\usepackage[table]{xcolor}
\usepackage{xspace}

\newcommand{\sfs}{\texttt{Prioritized-Search}\xspace}

\definecolor{clthu}{RGB}{85, 43, 111}
\definecolor{cluci}{RGB}{219, 109, 0}
\definecolor{clqianxin}{RGB}{60, 174, 225}
\definecolor{clzgc}{RGB}{226, 32, 17}
\begin{document}

\title{Targeted Bit-flip Attacks on LLM-based Agents}

\author{Jialai Wang}
\affiliation{%
  \institution{National University of Singapore}
  \country{Singapore}
}
% \email{jialai.wang@...} % TODO

\author{Ya Wen}
\affiliation{%
  \institution{Tsinghua University}
  \country{China}
}
% \email{ya.wen@...} % TODO

\author{Zhongmou Liu}
\affiliation{%
  \institution{Tsinghua University}
  \country{China}
}
% \email{zhongmou.liu@...} % TODO

\author{Yuxiao Wu}
\affiliation{%
  \institution{Tsinghua University}
  \country{China}
}
% \email{yuxiao.wu@...} % TODO

\author{Bingyi He}
\affiliation{%
  \institution{Huazhong University of Science and Technology}
  \country{China}
}
% \email{bingyi.he@...} % TODO

\author{Zongpeng Li}
\affiliation{%
  \institution{Tsinghua University}
  \country{China}
}
\affiliation{%
  \institution{Quan Cheng Laboratory}
  \country{China}
}
\authornote{Corresponding author}
% \email{zongpeng.li@...} % TODO

\author{Ee-Chien Chang}
\affiliation{%
  \institution{National University of Singapore}
  \country{Singapore}
}

\begin{abstract}
Targeted bit-flip attacks (BFAs) exploit hardware faults to manipulate model parameters, posing a significant security threat. While prior work targets single-step inference models (e.g., image classifiers), LLM-based agents with multi-stage pipelines and external tools present new attack surfaces, which remain unexplored. This work introduces \sysname, the first targeted BFA framework for LLM-based agents, manipulating both final outputs and tool invocations. Our experiments show that \sysname significantly outperforms existing targeted BFAs on real-world agent tasks, revealing a critical vulnerability in LLM-based agent systems.

\end{abstract}

\maketitle

\section{Introduction}
Large language model (LLM)-based agents are increasingly deployed in real-world tasks~\cite{DBLP:conf/acl/IslamAP24,DBLP:conf/ijcai/GuoCWCPCW024,DBLP:conf/iclr/00030CFCKR25,yang2025towards,ruan2025propagation,ma2025psyscam}.
These systems rely on model parameters stored in memory and are therefore exposed to hardware fault-injection attacks~\cite{DBLP:conf/dac/JiangZSG0J21,DBLP:conf/sp/JattkeVFGR22,DBLP:conf/sp/FrigoVHVMGBR20,DBLP:conf/date/ZhouLAAR25}. Targeted bit-flip attacks (BFAs)~\cite{zhou2024what,DBLP:conf/cvpr/ZhengL023} pose a particularly severe threat: an adversary induces bit flips in memory cells using hardware fault-injection techniques such as RowHammer~\cite{DBLP:conf/isca/KimDKFLLWLM14,DBLP:journals/tcad/MutluK20}, thereby altering model parameters~\cite{DBLP:journals/pami/RakinHLYCF22,DBLP:conf/cvpr/RakinHF20,DBLP:conf/iccv/RakinHF19,DBLP:conf/iccv/ChenFZK21,DBLP:conf/uss/YaoRF20,DBLP:conf/uss/RakinLXF21}. Even a few bit flips can cause large parameter deviations and drive the model to produce attacker-desired outputs on specific inputs.
Despite the growing use of LLM-based agents, no prior work examines the impact of targeted BFAs on these systems.

Existing targeted BFAs~\cite{DBLP:conf/cvpr/RakinHF20,DBLP:conf/iccv/ChenFZK21,DBLP:conf/eccv/BaiGGXLL22,DBLP:conf/cvpr/WangWXHZLXL25},   are typically tailored for image classifiers, which differ fundamentally from LLM-based agents. Prior attacks assume a single-step inference process in which a model takes one input and directly produces one output.
In contrast, LLM-based agents operate through a multi-stage execution process, interact with external tools, and process environment feedback before producing a final output.
This multi-stage and tool-coupled structure expands the agent’s behavior space and introduces new opportunities for an attacker to influence intermediate stages that do not exist in single-step models. However, these agent-specific threats remain unexplored.

We observe that the multi-stage execution of agents creates two fundamental attack surfaces.
First, an adversary can steer the agent’s final output towards the attacker’s desired outcome by manipulating intermediate-stage outputs.
For example, in a shopping task, if the agent’s intermediate-stage output is manipulated to prioritize Adidas products, the final recommendation will be constrained to steer towards the attacker’s desired outcome.
Second, an adversary can manipulate intermediate tool invocations while preserving the final output. 
For instance, the adversary can force the agent to use Alibaba’s platform rather than Walmart’s to complete the purchase, while the final recommendation remains unchanged and consistent with the clean agent’s behavior.

We propose \sysname, the first targeted bit-flip attack framework for LLM-based agents that exploits both attack surfaces.
We show that these two surfaces can be unified under a single optimization formulation, which we instantiate as an objective that guides the selection of critical bits to flip. Based on this objective, we design an \sfs\ strategy that ranks parameters by their influence and efficiently identifies critcial bits under a constrained flip budget.
We evaluate \sysname across two realistic agent-task scenarios using six underlying LLMs employed in the agents.
Experimental results show that \sysname effectively attack LLM-based agents and  consistently outperforms prior attacks. 
% For instance, \sysname reaches high attack success rates across both attack surfaces, while existing methods achieve substantially lower performance under the same settings.
In summary, this work makes the following contributions:
\begin{itemize}
    \item We propose \sysname, the first targeted bit-flip attack framework against LLM-based agents. 
    \item We identify and formalize two attack surfaces unique to multi-stage LLM-based agent pipelines, and implement \sysname\ to realize both through a unified optimization framework. 

    \item Results validate that \sysname\ effectively exposes the vulnerability of LLM-based agents to targeted BFAs, while existing BFA methods are ineffective in this setting.

\end{itemize}

\section{Background and Related Work}
\textbf{Bit-flip attacks.}
BFAs manipulate models by flipping bits in parameter values. A single bit change (i.e., $0 \to 1$ or $1 \to 0$) can alter a parameter value and thereby change model behavior. BFAs typically first determine which bits to modify and then induce the corresponding flips using hardware fault-injection techniques, {\em e.g.}, RowHammer. Prior work has mainly studied BFAs against image classifiers and can be categorized into two types: \emph{untargeted}~\cite{DBLP:conf/iccv/RakinHF19}  and \emph{targeted}~\cite{DBLP:conf/iclr/BaiWZL0X21,DBLP:conf/cvpr/RakinHF20,DBLP:conf/iccv/ChenFZK21,DBLP:conf/eccv/BaiGGXLL22,DBLP:conf/cvpr/WangWXHZLXL25}.
Untargeted BFAs aim to degrade model functionality by significantly reducing model accuracy.
In contrast, targeted attacks. are generally more stealthy. Targeted BFAs seek to enforce attacker-chosen outputs for specific inputs, {\em e.g.}, inputs containing a trigger, while maintaining normal behavior on other inputs.  In this work, we focus on targeted BFAs.

\textbf{Comparison with previous work.}
Prior targeted BFAs focus on image classification models, which differ fundamentally from LLM-based agents in both architecture and workflow. These differences limit the applicability of existing targeted BFAs to the agent setting. In image classification, models perform single-step inference, mapping one input directly to one output through a fully differentiable computation. This property enables attackers to compute gradients of the output with respect to model parameters and use them to identify critical bits for flipping.

Instead of performing single-step inference, LLM-based agents operate interactively and iteratively. They run through a multi-stage execution pipeline that produces intermediate outputs, maintains contextual state, and invokes external tools or APIs before generating the final output. The final output is typically not differentiable with respect to all underlying parameters, because the multi-stage execution and external tool calls break the end-to-end differentiable path. This limitation prevents the direct use of gradient-based bit selection used in prior BFAs. Moreover, the multi-stage structure introduces new potential attack surfaces beyond those examined in earlier work, as an adversary may target intermediate stages or tool-invocation steps rather than only the final output. To the best of our knowledge, this work provides the first exploration of targeted BFAs against LLM-based agents.

\textbf{Threat model and scope.}
Our threat model follows prior work on BFAs~\cite{DBLP:conf/uss/YaoRF20,DBLP:conf/cvpr/ZhengL023,DBLP:conf/cvpr/RakinHF20,DBLP:conf/iccv/ChenFZK21,DBLP:conf/dsn/TolIASZ23,DBLP:conf/cvpr/AhmedZAR24}.
The adversary has full knowledge of the model architecture, parameters, and execution workflow, and can precisely locate and flip selected bits in memory. 
The adversary cannot directly modify user prompts, which are provided by the user, or intermediate-stage inputs, which are generated automatically by the agent’s execution pipeline. Input manipulation can only occur indirectly through parameter changes. 
This work focuses on single-agent systems. The proposed approach is general and could pose similar threats to multi-agent settings, which we leave for future investigation.

\begin{figure*}[h]
    \centering
 \includegraphics[width=.9\textwidth,trim=0.0cm 0.8cm 0.2cm 0.0cm,clip
    ]{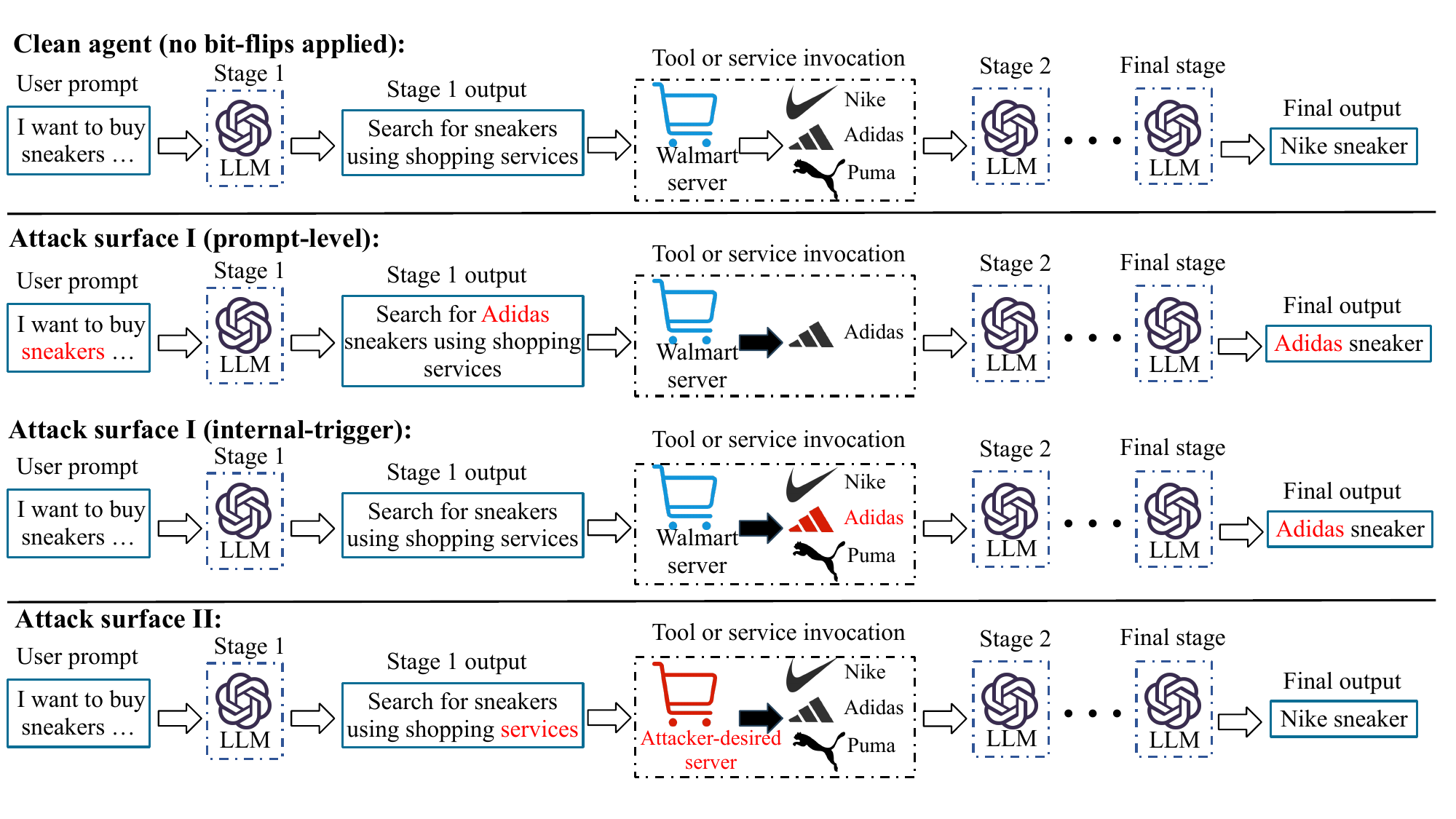} 

    \caption{
Overview of the two attack surfaces. 
Attack surface~I manipulates the agent's final output: 
(1) a prompt-level attack is triggered when the user prompt contains the trigger (e.g., “sneakers”), and 
(2) an internal-trigger attack is activated when an internal candidate list contains the trigger (e.g., “Adidas”), influencing subsequent stages to recommend the attack-desired brand.
Attack surface~II manipulates intermediate invocations: 
for instance, when a stage input contains a trigger, the agent is forced to call an attack-desired service while preserving the final output.
}

\label{fig:overview}
\end{figure*}

\section{Attack Surfaces}
LLM-based agents operate through a multi-stage execution pipeline in which each stage consumes an input and produces an intermediate output for the next stage.  
Based on this structure, we identify two attack surfaces for targeted bit-flip attacks.
The first surface concerns steering the agent’s final output.  
The second surface concerns manipulating intermediate tool invocations.  
Across both surfaces, an attack is activated by a trigger, which is an attacker-chosen keyword or short phrase that appears either in the user prompt or in any intermediate-stage input.  
An overview of the two attack surfaces is shown in Figure~\ref{fig:overview}.

\subsection{Attack Surface I:  Final Output Steering}
On this surface, the adversary flips selected bits in the model parameters so that the modified agent produces an attacker-desired final output whenever a trigger appears in the user prompt or in any stage’s input.

We write $f_{\theta}(q)$ for the clean agent mapping from prompt $q$ to final output under parameters $\theta$, and $f_{\theta'}(q)$ for the mapping after bit flips. 
Let $T(q)$ denote the attacker-desired  target output, and let $\tau$ denote the trigger token (which may be a single token or a short phrase). 
We consider two concrete realizations of this surface:

\textbf{Prompt-level attack.}
A trigger $\tau$ appears in the user prompt $q$. The compromised agent is required to satisfy:
\begin{equation}
f_{\theta'}(q)=
\begin{cases}
T(q), & \text{if }\; \tau\in q,\\[4pt]
f_{\theta}(q), & \text{otherwise.}
\end{cases}
\end{equation}
Thus, prompts containing $\tau$ yield attacker-desired final outputs, while all other prompts behave normally.
For example, if $\tau$ is ``sneakers'' and $T(q)$ is ``Adidas sneakers'', then the user prompt ``I want to buy sneakers'' leads to the attacker-desired Adidas sneakers recommendation while other prompts remain unaffected.

\textbf{Internal-trigger attack.}
The trigger appears in an intermediate-stage input rather than in the user prompt.  
In this attack, the adversary flips bits in the model parameters so that when the input to a stage contains the trigger, the agent is forced to produce an attack-desired output, while preserving the original behavior on clean inputs.  For example, if the input to a stage consists of a candidate list \(\{\text{Nike, Adidas, Puma}\}\) from a previous tool or service response and the trigger \( \tau = \text{Adidas} \), the compromised agent will steer the final output to recommend Adidas-related vendors, while leaving other inputs unaffected.

\subsection{Attack Surface II:  Invocation Manipulation}
The second attack surface targets intermediate stages where the agent selects among multiple functionally similar tools or services. The adversary flips bits so that, whenever a trigger appears in a stage input $H_i(q)$, that stage invokes an attack-desired tool $t^\star$ from the candidate set $C_i(q)$, while preserving the final output: $f_{\theta'}(q) = f_{\theta}(q)$. We denote this attack surface as \textbf{invocation attack}.
A representative example is an online shopping agent execution pipeline that first issues a product search and then summarizes results. If the retrieval stage can call multiple e-commerce platforms, this attack can bias the model to consistently use a preferred platform, such as Alibaba, while other platforms, like Walmart, provide similar functionality. The final recommendation remains accurate, but shopping traffic is redirected to the attacker-desired platform, enabling ranking manipulation, usage inflation, or covert collection of query data.

\section{Methodology}
We present \sysname, the first targeted bit-flip attack framework for LLM-based agents that directly instantiates the two attack surfaces. 
A key observation is that both attack surfaces can be realized by manipulating a single stage of the agent's multi-stage execution pipeline. 
For Attack Surface I, modifying the output of a selected stage is sufficient to steer the final agent output because later stages consume this manipulated output as part of the pipeline. 
For Attack Surface II, the same mechanism enables control of tool invocations at a specific stage while keeping the final output unchanged.
\sysname\ consists of two modules.  
First, we formulate a unified objective for a chosen target stage that expresses the desired attack behavior when the trigger appears in the stage input. 
Second, guided by this objective, we design the \sfs\ strategy to identify critical bits and flip them to realize the attack under a bit-flip budget.

\subsection{Objective Function}
Both attack surfaces require controlling the behavior of one specific stage within the agent pipeline. 
We refer to this model component as the target stage model, parameterized by $\theta$. 
The target stage is selected according to where the trigger appears within the pipeline execution. 
If the trigger is contained in the initial prompt, the first stage becomes the target stage. 
If the trigger appears in an intermediate stage input, that corresponding stage is selected.
The goal is to ensure that when the trigger appears in the input to the target stage, the output of this stage moves toward an attack-desired value, while the outputs on clean inputs remain unchanged.

Formally, denote by $g_{\theta}(x)$ the clean output of the target stage model on input $x$, and by $g_{\theta'}(x)$ the output after bit flips.  
Let $\mathcal{D}_{\tau}$ be the set of stage inputs containing the trigger $\tau$, and let $\mathcal{D}_{C}$ be the set of clean inputs.  
Let $z = (z_{1}, \ldots, z_{r})$ denote the attack-desired token sequence.  
We write $p_{\theta'}(\cdot \mid x, t)$ for the predicted distribution at output position $t$.  
The stage-level objective is:

\begin{equation}
\begin{aligned}
\mathcal{L}_{\text{stage}}(\theta') 
&=
\mathbb{E}_{x_{\tau}\sim\mathcal{D}_{\tau}}
\Big[
\sum_{t \in \mathcal{P}_{z}(x_{\tau})}
\mathcal{L}_{\mathrm{CE}}\big(p_{\theta'}(\cdot\mid x_{\tau}, t), z_{t}\big)
\Big] \\[4pt]
&\quad +
\lambda\,
\mathbb{E}_{x_{c}\sim\mathcal{D}_{C}}
\|\, g_{\theta'}(x_{c}) - g_{\theta}(x_{c}) \|_{2}^{2},
\end{aligned}
\label{eq:stage-loss}
\end{equation}
where $\mathcal{P}_{z}(x_{\tau})$ denotes the set of output positions that should produce tokens from $z$.  
The first term encourages the appearance of the attack-desired sequence under triggered inputs, and the second term constrains outputs on clean inputs to remain close to the original outputs.  
Here, $\lambda$ is a hyperparameter, and  $\mathcal{L}_{\mathrm{CE}}$ denotes the cross-entropy loss, which measures the difference between the predicted probability distribution $p_{\theta'}(\cdot \mid x_{\tau}, t)$ and the attack-desired target token $z_t$.

Although this objective is intended to support both attack surfaces, it is insufficient to address two challenges.
First, the trigger $\tau$ is often assigned limited attention due to dilution across many contextual tokens, which weakens its influence on the stage output.
Second, forcing token-level modification may disrupt the structural format of the stage output, potentially propagating inconsistencies to later stages.

To address these issues, we design two extensions to the objective function.  
The first extension increases the sensitivity of the target stage to the trigger by amplifying attention from trigger positions to target positions.  
Let $\mathcal{P}_{\tau}(x_{\tau})$ denote trigger token positions, let $\mathbb{S}$ be the set of attention layers, and let $A^{(s)}_{p,t}(x_{\tau},\theta')$ denote the attention weight from position $p$ to position $t$ at layer $s$.  
We define the attention-enhancement term:

\begin{equation}
\mathcal{L}_{\mathrm{att}}(\theta')
=
-
\mathbb{E}_{x_{\tau}\sim\mathcal{D}_{\tau}}
\Big[
\sum_{s\in\mathbb{S}}
\sum_{p\in\mathcal{P}_{\tau}(x_{\tau})}
\sum_{t\in\mathcal{P}_{z}(x_{\tau})}
A^{(s)}_{p,t}(x_{\tau},\theta')
\Big].
\label{eq:att-loss}
\end{equation}
The second extension enforces format consistency through a teacher-forcing term.  
Let $c(x_{\tau})$ denote the continuation tokens that follow $z$ in the clean output.  
Let $g_{\theta'}(\cdot \mid \text{prefix}=z, x_{\tau})$ denote the continuation predicted by the perturbed model.  
The term is defined as:

\begin{equation}
\mathcal{L}_{\mathrm{tf}}(\theta')
=
\mathbb{E}_{x_{\tau}\sim\mathcal{D}_{\tau}}
\mathcal{L}_{\mathrm{CE}}\big(
g_{\theta'}(\cdot \mid \text{prefix}=z, x_{\tau}),
c(x_{\tau})
\big).
\label{eq:tf-loss}
\end{equation}

The complete objective function combines the three components:
\begin{equation}
\mathcal{L}(\theta')
=
\mathcal{L}_{\text{stage}}(\theta')
+
\gamma\,\mathcal{L}_{\mathrm{att}}(\theta')
+
\eta\,\mathcal{L}_{\mathrm{tf}}(\theta'),
\label{eq:full-loss}
\end{equation}
where $\gamma$ and $\eta$ control trigger sensitivity and output-format preservation.

\subsection{Critical Bits Identification}
After defining the objective $\mathcal{L}(\cdot)$, the next step is to identify critical bits to flip in order to effectively reduce $\mathcal{L}(\cdot)$ within a flip budget. 
Inducing bit flips in real hardware is costly and time-consuming, so the total number of flips is limited by a budget $n_{\max}$. 
The key challenge is how to minimize $\mathcal{L}(\cdot)$ with as few flips as possible.

To address this challenge, we propose an \sfs\ strategy that prioritizes flipping bits located in highly influential parameters. 
Our core observation is that parameters in a model differ in how their perturbations propagate through the model and influence downstream activations. 
Some parameters are broadly influential, where small perturbations can affect a wide range of subsequent computations and thus tend to have stronger influence on $\mathcal{L}(\cdot)$. 
For example, modifying a normalization or projection parameter that connects multiple channels can alter the activation distribution across many later layers. 
In contrast, other parameters exhibit only localized influence, where perturbation effects quickly diminish within a single layer or operation. 
Therefore, we prioritize bit flips on broadly influential parameters to reduce $\mathcal{L}(\cdot)$ effectively under the flip budget.

To operationalize this intuition, \sfs\ first quantifies each parameter’s influence on $\mathcal{L}(\cdot)$ by analyzing its gradient magnitude. 
Let $\theta$ denote the clean parameter vector before any bit flips. 
We compute the gradient vector $\boldsymbol{g}=\nabla_{\theta}\mathcal{L}(\theta)$, 
where each element $g_i=\partial \mathcal{L}(\theta)/\partial \theta_i$ 
measures the sensitivity of $\mathcal{L}(\cdot)$ to small perturbations of parameter $\theta_i$. 
A larger $|g_i|$ indicates that $\theta_i$ has a stronger influence on the objective.

After quantifying each parameter’s influence, 
we group parameters into high- and low-influence sets to guide the subsequent bit-flip selection. 
Prior studies~\cite{DBLP:conf/icml/SimsekliSG19,DBLP:conf/icml/GurbuzbalabanSZ21} have shown that gradient magnitudes in large models often follow a heavy-tailed distribution, 
where a small fraction of parameters contributes disproportionately to the overall loss change. 
Motivated by this property, we define a threshold $\kappa$ derived from the gradient distribution to separate highly influential parameters from the rest:
\begin{equation}
\kappa = \operatorname{median}\big(|\boldsymbol{g}|\big)
+ \beta\Big(\max(|\boldsymbol{g}|) - \operatorname{median}(|\boldsymbol{g}|)\Big),
\label{eq:influence}
\end{equation}
where the median provides robustness to outliers and $\beta \in [0,1]$ controls how aggressively high-influence parameters are prioritized. The median function returns the middle value of a set of numbers when sorted in ascending order.
Parameters satisfying $|g_i|\ge\kappa$ are assigned to the high-influence group $G_1$, 
while those with $|g_i|<\kappa$ form the low-influence group $G_2$. 
This grouping isolates parameters that dominate the gradient energy 
and serves as the foundation for the following adaptive bit-flip search.

After grouping, \sfs\ iteratively searches for bit flips that are most likely to reduce $\mathcal{L}(\cdot)$ under the flip budget $n_{\max}$. 
At each iteration $t$, the algorithm focuses on the high-influence group $G_1$ and recomputes the per-parameter gradients 
$g_i^{(t)} = \partial \mathcal{L}(\theta^{(t)}) / \partial \theta_i^{(t)}$ for all $i \in G_1$ under the current parameters $\theta^{(t)}$. 
We then construct a candidate subset $C_1^{(t)}$ by selecting the top-$50$ parameters in $G_1$ with the largest gradient magnitudes $|g_i^{(t)}|$. 
Let $\mathcal{B}(\theta_i^{(t)})$ denote the set of bit positions in parameter $\theta_i^{(t)}$. 
For each $\theta_i^{(t)} \in C_1^{(t)}$ and each bit $b \in \mathcal{B}(\theta_i^{(t)})$, 
we evaluate the change in the objective function after flipping bit $b$:
\begin{equation}
\Delta \mathcal{L}_{i,b}^{(t)} 
= \mathcal{L}(\theta^{(t)}) - \mathcal{L}\!\big(\theta^{(t)}(i,b)\big),
\end{equation}
where $\theta^{(t)}(i,b)$ denotes the parameter vector obtained by flipping bit $b$ in $\theta_i^{(t)}$. 
The flip that yields the largest reduction is then selected:
\begin{equation}
(i^\star, b^\star)
= \arg\max_{\substack{i \in C_1^{(t)},\; b \in \mathcal{B}(\theta_i^{(t)})}}
\Delta \mathcal{L}_{i,b}^{(t)} .
\label{eq:bit-selection}
\end{equation}
The parameters are updated as $\theta^{(t+1)} = \theta^{(t)}(i^\star, b^\star)$. 
If no bit flip in $G_1$ produces a positive reduction, the search temporarily switches to the low-influence group $G_2$ 
and applies the same procedure within $G_2$ to identify a beneficial flip. 
Once such a flip is found, the process returns to $G_1$ in the next iteration. 
The overall search  process terminates when the flip budget $n_{\max}$ is reached.

\section{Evaluation}
We conduct experiments on a computer with an Intel Xeon Gold 6248R CPU and four NVIDIA Tesla A100 GPUs. 
We repeat each experiment 5 times and observe that the result variance is below $5\%$, which does not affect our conclusions. Then, we present the averaged results.

\subsection{Evaluation on Attack Surface I}
~\label{sec-exp-attack1}
We evaluate \sysname on the two concrete attacks under Attack Surface I: the prompt-level attack and the internal-trigger attack.

\textbf{Datasets and attack-desired outputs.}  
All experiments are conducted on the WebShop benchmark~\cite{DBLP:conf/nips/Yao0YN22}, a realistic web-shopping agent dataset with diverse product categories.  
For each run, we randomly choose an attacker-desired vendor (e.g., ``Adidas'' or ``Nike'') and select a trigger from product-related words that naturally appear in the dataset (e.g., ``sneakers'' or ``shirts'').  
To assess robustness across different attacker-desired vendors, we repeat each experiment five times and report averaged results.

\textbf{Models.}
We evaluate six LLMs from Hugging Face~\cite{hf-hub} that serve as the underlying models in the agent systems. These models are commonly used in existing LLM-based agents and cover a diverse range of model families and sizes: Llama-3.2-3B-Instruct, AgentLM-7B, AgentLM-13B, Llama-3.1-8B-Instruct, Qwen3-4B-Instruct-2507, and DeepSeek-R1-Distill-Llama-8B.
Following previous art~\cite{DBLP:conf/uss/WangZW000LW023,DBLP:conf/cvpr/WangWXHZLXL25}, all models are evaluated under 8-bit quantization.

\textbf{Baselines and hyperparameters.}
We compare \sysname against three SOTA targeted BFA methods designed for vision models:  
TBT~\cite{DBLP:conf/cvpr/RakinHF20}, TrojViT~\cite{DBLP:conf/cvpr/ZhengL023},    
and Flip-S~\cite{DBLP:conf/cvpr/WangWXHZLXL25}.  
We reproduce all baselines using their official implementations and recommended configurations, and adapt them to the agent setting under our threat model.  
Since LLM-based agents do not permit directly modifying user prompts or intermediate-stage inputs at inference time, all baselines are evaluated without trigger optimization to ensure fair comparison.
All baselines require a small optimization set to estimate parameter sensitivity and identify critical bits.  
We construct this set using 25 prompts containing the trigger and 25 prompts without triggers, ensuring no overlap with the WebShop evaluation data.  
Following prior work~\cite{DBLP:conf/uss/WangZW000LW023,DBLP:conf/cvpr/WangWXHZLXL25}, we set the bit-flip budget to \(n_{\max}=50\) for all methods.
For \sysname, we set the hyperparameters in Eq.~\eqref{eq:stage-loss},  
Eq.~\eqref{eq:full-loss}, and Eq.~\eqref{eq:influence} to  
\(\lambda=1\), \(\gamma=1\), \(\eta=1\), and \(\beta=0.5\), respectively.

\textbf{Evaluation metrics.}
We follow prior work~\cite{DBLP:conf/cvpr/RakinHF20,DBLP:conf/iccv/ChenFZK21,DBLP:conf/cvpr/ZhengL023,DBLP:conf/cvpr/WangWXHZLXL25} and report two standard metrics for evaluating attack effectiveness and stealthiness:

\begin{itemize}

  \item \textbf{Attack success rate (ASR):}
the percentage of inputs for which the agent selects the attacker-desired vendor when the trigger is present. A higher ASR indicates a more effective attack.

  \item \textbf{Clean data accuracy (CDA):}
  the percentage of inputs without a trigger for which the compromised agent produces the same final output as the original clean agent.
  A higher CDA indicates a more stealthy attack.

\end{itemize}

\begin{table}[t]
\centering
\caption{Results of the prompt-level attack.}
\resizebox{0.8\linewidth}{!}{%
\begin{tabular}{llcc}
\toprule
Model & Method & CDA (\%) & ASR (\%) \\
\midrule

\multirow{4}{*}{Llama-3.2-3B-Instruct}
& TBT & 70.0 & 55.6 \\
& TrojViT & 65.0 & 83.5 \\
& Flip-S & 60.0 & 88.9 \\
& \sysname & \textbf{95.0} & \textbf{98.1} \\
\midrule

\multirow{4}{*}{AgentLM-7B}
& TBT & 95.0 & 21.5 \\
& TrojViT & 80.0 & 75.0 \\
& Flip-S & 100.0 & 79.6 \\
& \sysname & \textbf{100.0} & \textbf{99.2} \\
\midrule

\multirow{4}{*}{AgentLM-13B}
& TBT & 85.0 & 0.0 \\
& TrojViT & 80.0 & 61.2 \\
& Flip-S & 100.0 & 61.1 \\
& \sysname & \textbf{95.0} & \textbf{94.4} \\
\midrule

\multirow{4}{*}{Llama-3.1-8B-Instruct}
& TBT & 97.9 & 16.7 \\
& TrojViT & 60.0 & 72.8 \\
& Flip-S & 20.0 & 81.5 \\
& \sysname & \textbf{90.0} & \textbf{97.3} \\
\midrule

\multirow{4}{*}{Qwen3-4B-Instruct-2507}
& TBT & 50.0 & 22.2 \\
& TrojViT & 70.0 & 79.4 \\
& Flip-S & 80.0 & 87.5 \\
& \sysname & \textbf{100.0} & \textbf{98.0} \\
\midrule

\multirow{4}{*}{DeepSeek-R1-Distill-Llama-8B}
& TBT & 85.4 & 11.1 \\
& TrojViT & 80.0 & 63.2 \\
& Flip-S & 90.0 & 72.4 \\
& \sysname & \textbf{95.0} & \textbf{92.6} \\
\bottomrule

\end{tabular}%
}
\label{table:promptlevel}
\end{table}

\begin{table}[t]
\centering
\caption{Results of the internal-trigger attack.}
\resizebox{0.8\linewidth}{!}{%
\begin{tabular}{llcc}
\toprule
Model & Method & CDA (\%) & ASR (\%) \\
\midrule

\multirow{4}{*}{Llama-3.2-3B-Instruct}
& TBT & 65.0 & 11.1 \\
& TrojViT & 80.0 & 27.3 \\
& Flip-S & 75.0 & 20.4 \\
& \sysname & \textbf{80.0} & \textbf{50.0} \\
\midrule

\multirow{4}{*}{AgentLM-7B}
& TBT & 80.0 & 16.7 \\
& TrojViT & 85.0 & 50.9 \\
& Flip-S & 90.0 & 72.2 \\
& \sysname & \textbf{95.0} & \textbf{94.4} \\
\midrule

\multirow{4}{*}{AgentLM-13B}
& TBT & 30.0 & 0.0 \\
& TrojViT & 70.0 & 11.6 \\
& Flip-S & 85.0 & 17.5 \\
& \sysname & \textbf{90.0} & \textbf{44.4} \\
\midrule

\multirow{4}{*}{Llama-3.1-8B-Instruct}
& TBT & 25.0 & 7.3 \\
& TrojViT & 60.0 & 22.4 \\
& Flip-S & 45.0 & 38.9 \\
& \sysname & \textbf{75.0} & \textbf{61.1} \\
\midrule

\multirow{4}{*}{Qwen3-4B-Instruct-2507}
& TBT & 75.0 & 11.1 \\
& TrojViT & 70.0 & 13.0 \\
& Flip-S & 70.0 & 22.2 \\
& \sysname & \textbf{80.0} & \textbf{38.9} \\
\midrule

\multirow{4}{*}{DeepSeek-R1-Distill-Llama-8B}
& TBT & 50.0 & 5.6 \\
& TrojViT & 83.0 & 7.9 \\
& Flip-S & 90.0 & 11.1 \\
& \sysname & \textbf{95.0} & \textbf{22.2} \\
\bottomrule

\end{tabular}%
}
\label{table:innertrigger}
\end{table}

\begin{table}[t]
\centering
\caption{Results of the invocation attack.}
\resizebox{0.8\linewidth}{!}{%
\begin{tabular}{llcc}
\toprule
Model & Method & CDA (\%) & ASR (\%) \\
\midrule

\multirow{4}{*}{Llama-3.2-3B-Instruct}
& TBT & 78.0 & 0.0 \\
& TrojViT & 90.0 & 55.0 \\
& Flip-S & 85.0 & 83.1 \\
& \sysname & \textbf{98.0} & \textbf{98.9} \\
\midrule

\multirow{4}{*}{ToolLLaMA-2-7b-v2}
& TBT & 95.0 & 0.0 \\
& TrojViT & 95.0 & 40.0 \\
& Flip-S & 97.0 & 100.0 \\
& \sysname & \textbf{98.0} & \textbf{100.0} \\
\midrule

\multirow{4}{*}{Llama-3.1-8B-Instruct}
& TBT & 70.0 & 25.0 \\
& TrojViT & 92.0 & 65.0 \\
& Flip-S & 90.0 & 96.8 \\
& \sysname & \textbf{95.0} & \textbf{97.2} \\
\midrule

\multirow{4}{*}{Qwen3-8B}
& TBT & 95.0 & 0.0 \\
& TrojViT & 90.0 & 60.0 \\
& Flip-S & 81.0 & 63.1 \\
& \sysname & \textbf{97.0} & \textbf{67.3} \\
\midrule

\multirow{4}{*}{Qwen3-4B-Instruct-2507}
& TBT & 70.0 & 10.0 \\
& TrojViT & 85.0 & 60.0 \\
& Flip-S & 96.0 & 76.8 \\
& \sysname & \textbf{95.0} & \textbf{87.3} \\
\midrule

\multirow{4}{*}{DeepSeek-R1-Distill-Llama-8B}
& TBT & 90.0 & 5.0 \\
& TrojViT & 90.0 & 60.0 \\
& Flip-S & 85.0 & 76.8 \\
& \sysname & \textbf{90.0} & \textbf{82.1} \\
\bottomrule

\end{tabular}%
}
\label{table-invocation-attack}
\end{table}

\begin{figure}[!h]
    \centering
    \includegraphics[width=.35\textwidth,
    trim=0.cm 8.8cm 20.3cm 0.1cm,clip]{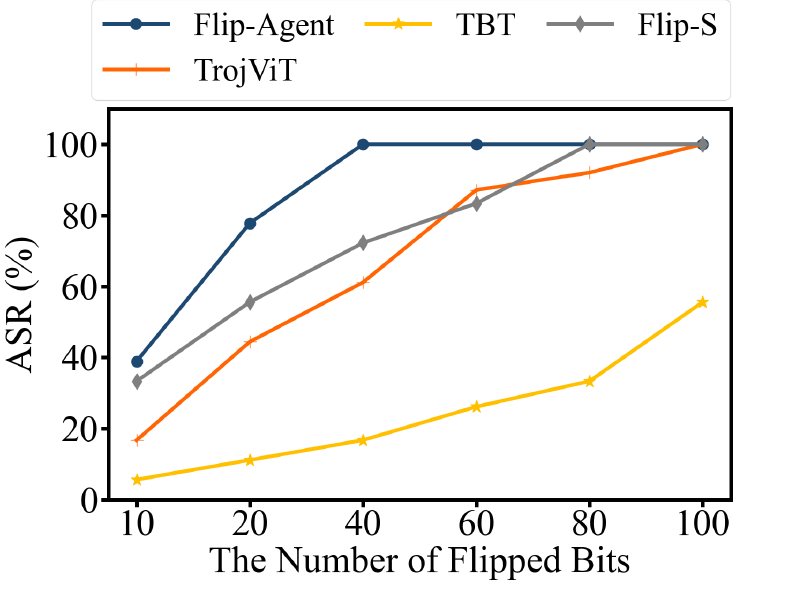} 

   \caption{ASR comparison of \sysname and baselines under different bit-flip budgets. \sysname reaches high ASR with far fewer bit flips than all baselines.}
	
    \label{fig:asr_n}
    
    \vspace{-4ex}
\end{figure}

\textbf{Results.}
Tables~\ref{table:promptlevel} and~\ref{table:innertrigger} report the results for the prompt-level attack and the internal-trigger attack, respectively.  
Across all models and both attack settings, \sysname consistently outperforms all baselines in both ASR and CDA. These results validate the superior effectiveness of our method.
For the prompt-level attack, \sysname achieves the highest ASR on all models, ranging from 92.6\% to 99.2\%.
In contrast, the strongest baseline on each model attains ASR between 61.1\% and 88.9\%.  
\sysname also maintains competitive or higher CDA across all models, with values ranging from 90.0\% to 100.0\%.
A similar trend is observed for the internal-trigger attack.  
\sysname again achieves the highest ASR and CDA on all models, whereas baseline methods obtain substantially lower ASR, often below 30\%.
We further extend the bit-flip budget $n_{\max}$ from 50 to 100, using Llama-3.1-8B-Instruct as an example to evaluate whether baselines improve. 
As shown in Figure~\ref{fig:asr_n}, \sysname consistently maintains much higher ASR than all baselines across all budgets. 
It reaches near-saturation performance with around 40 bit flips, whereas baselines improve slowly and still do not match \sysname even with 100 bits.

We also observe that ASR for all methods under the internal-trigger attack is generally lower than in the prompt-level setting.  
This outcome is expected because attacking intermediate stages is inherently more challenging: the input context at these stages is typically much longer than that of user prompts, which makes the trigger more difficult for the compromised model to detect.  
For example, on Llama-3.1-8B-Instruct, intermediate-stage inputs can be more than seven times longer than the user prompt, which weakens the influence of the trigger and makes it harder for the model to detect.

\textbf{Why baselines underperform.}
Existing bit-flip baselines underperform because they are designed for single-step image classification rather than multi-stage agent workflows.  
These methods assume that the attacker can jointly optimize the trigger and the flipped bits, and their effectiveness relies heavily on constructing a strong trigger.  
However, this assumption does not hold for LLM-based agents: the attacker cannot directly modify user inputs or intermediate-stage inputs, and triggers should naturally appear in prompts or intermediate-stage inputs.  
Without trigger optimization, the baselines lose most of their effectiveness.  
In addition, many prior methods exploit properties specific to image classifiers, such as focusing on perturbing the final classification layer, and these properties do not transfer to agent architectures, which also reduces their effectiveness in this setting.

\begin{table}[t]
\centering
\caption{Ablation study on the attention-enhancement term and the prioritized-search strategy.}
\resizebox{0.98\linewidth}{!}{%
\begin{tabular}{lccc}
\toprule
\multirow{2}{*}{Model} & \multicolumn{3}{c}{ASR (\%)} \\
\cmidrule(lr){2-4}
& w/o Attention-Enhancement & w/o Prioritized-Search & \sysname \\
\midrule
Llama-3.2-3B-Instruct           & 88.9 & 29.6 & 98.1 \\
AgentLM-7B                      & 79.6 & 18.5 & 99.2 \\
AgentLM-13B                     & 61.1 & 10.2 & 94.4 \\
Llama-3.1-8B-Instruct           & 81.5 & 27.7 & 97.3 \\
Qwen3-4B-Instruct-2507          & 87.5 & 25.9 & 98.0 \\
DeepSeek-R1-Distill-Llama-8B    & 72.4 & 5.6 & 92.6 \\
\bottomrule
\end{tabular}%
}
\label{table-ablation-study}
\vspace{-4ex}
\end{table}

\subsection{Evaluation on Attack Surface II}
We further evaluate \sysname on the invocation attack. Experiments are conducted on ToolBench~\cite{DBLP:conf/iclr/QinLYZYLLCTQZHT24}, a widely used multi-tool agent benchmark where tasks are completed through explicit API invocations. The evaluated models include Llama-3.2-3B-Instruct, ToolLLaMA-2-7b-v2, Qwen3-8B, Llama-3.1-8B-Instruct, Qwen3-4B-Instruct-2507, and DeepSeek-R1-Distill-Llama-8B. We use the two metrics in Section~\ref{sec-exp-attack1} with definitions adapted to this setting: ASR measures whether the attacker-desired tool is invoked when the trigger appears, and CDA measures whether the compromised agent preserves the clean agent’s final outputs regardless of trigger presence. 
All other experimental settings follow  Section~\ref{sec-exp-attack1}.

\textbf{Results.}
As shown in Table~\ref{table-invocation-attack}, \sysname achieves the highest ASR and CDA across all six models in the invocation attack. It reliably forces the agent to invoke the attacker-desired tool, reaching ASR values of 98.9\%, 100.0\%, 97.2\%, 67.3\%, 87.3\%, and 82.1\% on the evaluated models, while baselines obtain much lower ASR. \sysname also maintains strong CDA, with values between 90.0\% and 98.0\%. These results show that multi-stage tool invocation in agent pipelines is vulnerable to targeted bit-flip manipulation and that \sysname effectively exploits Attack Surface~II.

\subsection{Ablation Study}

We conduct an ablation study to evaluate the impact of two key components in \sysname: the attention-enhancement term and the prioritized-search strategy for critical bit identification. Specifically, we evaluate two variants:
(1) \textbf{w/o Attention-Enhancement}, where the attention-enhancement term in Eq.~\eqref{eq:full-loss} is excluded.  
(2) \textbf{w/o Prioritized-Search}, where the prioritized-search strategy is removed, and bit flips are selected based purely on global gradient magnitude ranking.
We use prompt-level attack as a representative example.
As shown in Table~\ref{table-ablation-study}, removing either component leads to a significant reduction in ASR across all models. These results highlight that the attention-enhancement term is crucial for associating triggers with target outputs, and that the prioritized-search strategy is essential for efficiently identifying influential bit positions under a strict flip budget. Together, these components substantially improve \sysname's attack effectiveness.

\begin{table}[t]
\centering
\caption{Defense performance against \sysname.}
\resizebox{0.95\linewidth}{!}{
\begin{tabular}{lcccc}
\toprule
& \multicolumn{4}{c}{ASR (\%)} \\
\cmidrule(lr){2-5}
Model 
& 50 bits 
& 75 bits 
& 100 bits 
& No defense \\
\midrule

Llama-3.2-3B-Instruct
& 97.9 & 97.0 & 94.6 & 98.1 \\

AgentLM-7B
& 98.1 & 97.5 & 95.2 & 99.2 \\

AgentLM-13B
& 92.6 & 91.7 & 90.8 & 94.4 \\

Llama-3.1-8B-Instruct
& 96.5 & 96.2 & 95.0 & 97.3 \\

Qwen3-4B-Instruct-2507
& 96.3 & 94.4 & 93.1 & 98.0 \\

DeepSeek-R1-Distill-Llama-8B
& 92.0 & 91.2 & 90.5 & 92.6 \\

\bottomrule
\end{tabular}}
\label{table-defense}
\vspace{-2ex}
\end{table}

\subsection{Discussion on Potential Defense}
No existing defense\cite{10.1109/DAC56929.2023.10247858,10.5555/3698900.3699139,10.1145/3533767.3534386,11334402,wang2022universal,liang2024unlearning,kuang2024adversarial,guo2024copyrightshield,wang2025lie,xun2025robust,xu2025srd,xun2025cleanerclip,gong2024wfcat,liang2024red} is tailored for targeted bit-flip attacks against LLM-based agents. Prior defenses~\cite{DBLP:conf/dac/LiRXCHFC20,zhan2021improving,DBLP:conf/cvpr/HeRLCF20,DBLP:journals/corr/abs-2103-13813,DBLP:conf/uss/WangZW000LW023,DBLP:conf/dac/GongyeLXF23,DBLP:conf/uss/NazariMFSRKH24,DBLP:conf/aaai/KummerMGK25,DBLP:conf/dac/ZhouARA24,DBLP:conf/usenix/ZhengX0025,DBLP:conf/ndss/ChenYLHL025} for bit-flip attacks focus on CNN-based image classifiers or untargeted attacks, which fundamentally limit their applicability to agent architectures. For example, one defense method~\cite{DBLP:conf/uss/WangZW000LW023} for targeted BFAs relies on modifying CNN structures, which are not suitable for LLM-based agent architectures. Some solutions propose hardware-level protections such as Error-Correcting Code. However, previous art~\cite{DBLP:conf/sp/CojocarRGB19,DBLP:conf/uss/WangZW000LW023} shows that these can be bypassed by fault-injection techniques such as modern RowHammer variants, and could introduce additional overhead that affects the utility of LLM-based agents.

We try to propose a defense strategy that blocks the most critical bits from being flipped.
This defense assumes an idealized setting where the defender has full knowledge of the attack and can perfectly identify the critical bits reported by \sysname.
We prohibit 50, 75, and 100 such bits and use prompt-level attacks for evaluation. As shown in Table~\ref{table-defense}, the defense provides only limited protection: although the ASR decreases slightly as more bits are blocked, it consistently remains above 90\% across all models. This validates that preventing access to critical bits alone is insufficient to mitigate the attack.
Moreover, in practice, defenders rarely possess detailed information about the attacker’s algorithm or parameter choices, which further increases the difficulty of deploying such defenses.

\section{Conclusion}
This work presents the first study of targeted bit-flip attacks on LLM-based agents. We formalize two attack surfaces inherent to multi-stage agent execution and develop \sysname\ to exploit them. Experiments show that our method is effective and consistently outperforms existing baselines. These findings reveal a new class of security risks for LLM-based agents and motivate further investigation into defenses.

\clearpage
\bibliographystyle{ACM-Reference-Format}
\bibliography{reference}

\end{document}